\documentclass[linenumbers,twocolumn,twocolappendix]{aastex631}
\nolinenumbers

\usepackage{tabularray}
\usepackage{graphicx}

\let\tablenum\undefined 
\usepackage{siunitx}

\usepackage{amsmath}
\usepackage{hyperref}
\UseTblrLibrary{booktabs}

\begin{document}
\title{A Suppressed Volumetric Rate of High-Luminosity Mid-Infrared Selected Tidal Disruption Events}

\author[0009-0009-1605-615X]{Prajna Nair}
\affiliation{MIT Kavli Institute for Astrophysics and Space Research, Massachusetts Institute of Technology, Cambridge, MA 02139, USA}

\author[0009-0001-9034-6261]{Christos Panagiotou}
\affiliation{MIT Kavli Institute for Astrophysics and Space Research, Massachusetts Institute of Technology, Cambridge, MA 02139, USA}

\author[0000-0003-4127-0739]{Megan Masterson}
\affiliation{MIT Kavli Institute for Astrophysics and Space Research, Massachusetts Institute of Technology, Cambridge, MA 02139, USA}

\author[0000-0002-8989-0542]{Kishalay De}
\affiliation{Department of Astronomy and Columbia Astrophysics Laboratory, Columbia University, 550 W 120th St. MC 5246, New York, NY 10027, USA}
\affiliation{Center for Computational Astrophysics, Flatiron Institute, 162 5th Ave., New York, NY 10010, USA}

\author[0000-0003-0172-0854]{Erin Kara}
\affiliation{MIT Kavli Institute for Astrophysics and Space Research, Massachusetts Institute of Technology, Cambridge, MA 02139, USA}

\author{Eleanor Winkler}
\affiliation{MIT Kavli Institute for Astrophysics and Space Research, Massachusetts Institute of Technology, Cambridge, MA 02139, USA}

\author{M. Subhi Abo Rdan}
\affiliation{MIT Kavli Institute for Astrophysics and Space Research, Massachusetts Institute of Technology, Cambridge, MA 02139, USA}

\correspondingauthor{Prajna Nair}
\email{prajna25@mit.edu}

\begin{abstract}

\nolinenumbers
Tidal Disruption Events (TDEs) serve as direct probes of the population of supermassive black holes in the center of galaxies and are nowadays regularly detected in optical wide-field time-domain sky surveys. Recent studies have demonstrated that a large fraction of TDEs can be uniquely identified in the infrared (IR) waveband, but these studies have to date been limited to relatively nearby events. In this work, we searched for highly luminous IR-bright TDEs that are rare and thus missed by searches in the local universe. We performed a systematic search of the NEOWISE archive and developed a new selection criterion based on the evolution of the W1-W2 color to select TDE candidates. We identified 10 IR bright TDEs with peak luminosities above $L_{\rm peak\, W2} \simeq 3 \times 10^{43}$ erg s$^{-1}$ and estimated an event rate of $1.2^{+0.5}_{-0.4}\times10^{-10}$ Mpc$^{-3}$year$^{-1}$ for the luminosity range of our sample. Compared to the existing local luminosity function of lower luminosity events, we detect a suppressed rate for these highly luminous events. This turn-over in the luminosity function can be naturally explained by the suppressed amount of TDEs taking place in systems with larger black hole masses, thereby confirming the TDE nature of our sources.

\end{abstract}

\section{Introduction} \label{sec:intro}

Tidal Disruption Events, or TDEs, are transient astrophysical phenomena that occur when a star’s orbit brings it within the tidal radius of a supermassive black hole, but still outside of its event horizon. The tidal forces of the black hole overcome the star’s self-gravity, causing the star to be ripped apart. A portion of the resultant stellar debris falls back into the black hole, forming an accretion disk around it, which leads to a flare of radiation that can be detected across the electromagnetic spectrum \citep{rees_tidal_1988, gezari_tidal_2021}.

The first TDE candidates were detected in the 1990s with the  ROSAT all-sky X-ray survey  \citep{komossa_discovery_1999, donley_large-amplitude_2002}. In the early 2010s, optical emission from TDEs was first detected, which opened up a new method of discovery \citep{van_velzen_optical_2011}. 
Later, optical surveys such as the All-Sky Automated Survey for Supernovae \citep[ASAS-SN,][]{shappee_man_2014}, the Asteroid Terrestrial-impact Last Alert System \citep[ATLAS,][]{tonry_atlas_2018}, and the Zwicky Transient Facility \citep[ZTF,][]{bellm_zwicky_2019} have led to the discovery of on the order of 100 TDEs \citep[][]{gezari_tidal_2021}.

This increase in the number of TDEs being found due to the optical time-domain surveys, as well as the advent of new X-rays surveys such as eROSITA \citep{predehl_erosita_2021}, has enabled detailed studies of the TDE population and demographics \citep[e.g.][]{sazonov_first_2021, hammerstein_final_2022, yao_tidal_2023,2026A&A...708A.374Z}. 

The results of these population studies suggest that TDEs preferentially occur in post-starburst galaxies \citep{arcavi_continuum_2014, french_tidal_2016}, and are possibly connected to galaxies that have undergone recent mergers \citep{hammerstein_tidal_2021, 2021MNRAS.500.3944P, 2024ApJ...969L..17W}. 
However, as noted by \cite{roth_forward_2021}, the lack of optical TDEs discovered in star-forming galaxies could be due to 
dust obscuration in those galaxies. It has been pointed out that TDEs obscured by dust in the optical wavelength range may be detectable in the infrared (IR) band. IR emission is thought to arise from the reprocessed optical, UV and X-ray emission from nuclear dust \citep{lu_infrared_2016} \textemdash a conjecture supported by the discovery of IR counterparts during follow-up observations of optical TDEs \citep[e.g.][]{velzen_discovery_2016, Jiang2016,2016ApJ...832..188D}. 

This motivated the use of IR surveys to search for dust-obscured TDEs missed by optical surveys. For instance, the Mid-InfraRed Outbursts in Nearby Galaxies survey \citep[MIRONG,][]{jiang_mid-infrared_2021-1} has found a number of mid-IR flares in low-redshift (z$<$0.35) SDSS galaxies that have no optical counterpart, many of which are TDE candidates \citep{2022ApJS..258...21W}. \cite{panagiotou_luminous_2023} reported the discovery of a dust-obscured TDE in the star-forming galaxy NGC 7392 at just 42 Mpc, indicating that there could be a sizable population of such TDEs yet to be studied.
\cite{masterson_new_2024} performed a systematic search for mid-IR bright TDEs in the nearby universe, out to a redshift of z$\lesssim$0.05, and found 12 TDEs, 
none of which were found to be in post-starburst systems. Only one of these 12 TDEs had a significant optical counterpart, pointing at a population of TDEs that can be uniquely identified in the IR waveband. Follow-up efforts have confirmed the TDE nature of these transients, as JWST MIRI spectroscopy revealed accretion-driven emission lines but different environments compared to active galaxies \citep{Masterson2025ApJ}. 

An important observable in TDE population studies is the existence of a suppression of events around high mass black holes, which arises because the tidal radius becomes smaller than the black hole's event horizon radius above a specific mass limit \citep[$M_{\rm BH} \sim 10^{8}$\(M_\odot\) for a Schwarzschild black hole and $M_{\rm BH} \sim 7\times10^{8}$\(M_\odot\) for maximally spinning black holes;][]{kesden_tidal-disruption_2012}. Assuming that the TDE flux scales with black hole mass, this also naturally leads to a reduced rate for highly luminous events. Since this mass limit depends on the spin of the black hole, the suppressed rate offers a unique probe of the spin distribution of quiescent supermassive black holes, which is otherwise challenging to investigate. The detection of such a suppression also serves as rather unambiguous evidence for the TDE nature of the observed flares \citep{Velzen2018}. In fact, a suppression of event rates in more luminous TDEs has already been detected in samples of optical \citep{yao_tidal_2023} and X-ray \citep{Guolo2024,grotova_population_2025} TDEs. 

The existing population of IR-selected TDEs from \cite{masterson_new_2024} only extended to $z \lesssim 0.05$, a small volume of the universe that thus misses the rare high-luminosity events. In order to obtain a complete census of the TDE population, we need to extend this work to the high-luminosity regime. In this work, we compile a sample of highly luminous IR-bright TDEs, which allows us to investigate a potential rate suppression at higher luminosities. 

This paper is organized as follows. We discuss the initial sorting algorithm and compilation of IR TDE candidates in Section \ref{selecting_sample}, including making selection cuts based on photometric and spectroscopic data. Section \ref{multi-wav} details our search for multi-wavelength counterparts associated with the IR flares. We calculate the IR TDE rate and discuss our results in Sections \ref{sec:lum_func} and \ref{sec:disc}, before summarizing our findings in Section \ref{sec:conclusion}.
Throughout this work, we assume a flat $\Lambda$CDM cosmology, with $H_0=67.66$ km s$^{-1}$Mpc$^{-1}$, $\Omega_M=0.311$ and $\Omega_\Lambda=0.689$.

\section{Sample Selection} 
\label{selecting_sample}

Our goal is to find a sample of highly luminous mid-IR TDEs. To this end, we performed a systematic search of the NEOWISE archive to identify large amplitude transients in time-resolved co-added images released as part of the unWISE project \citep{Lang2014, Meisner2018}. We employed a customized pipeline based on the ZOGY algorithm \citep{Zackay2016} to perform image subtraction on the NEOWISE images using the full-depth co-added images of the 2010-2011 WISE mission as reference images \citep[][De et al. in prep.]{De2023Natur}. We cross-matched the detected transients with the GLADE+ sample of galaxies, a catalog with high completeness out to redshift $z\sim 0.2$ \citep{dalya_glade_2022}. We classified our transients as nuclear if occurred within 2'' from the center of known galaxies and we then applied a series of tests, based on the photometry and spectra of these sources, in order to identify TDEs within this group of nuclear transients and exclude non-TDE contaminants, such as flaring AGN and supernovae. These tests are summarized in Table \ref{tab:selection} and are detailed in the sections below. It should be noted that while \cite{masterson_new_2024} employ a similar series of tests to ensure a clean sample of TDEs in their study, they cross-match WISE transients to the Census of the Local Universe compiled catalog, which only contains galaxies within $200$ Mpc, while our sources extend up to $1400$ Mpc. The larger volume we consider here will allow us to probe the more rare luminous sources.

\subsection{Photometric Analysis of TDE Candidates} 
\label{sec:photo_analysis}

First, we considered the shape of the transient's light curve and aimed to select sources that broadly resemble the light curve of known mid-IR TDEs \citep{masterson_new_2024}. More specifically, we selected transient events that last at least 2 years in total, feature a smooth increase to their peak flux, and a slower decrease to quiescence, while we excluded sources for which the maximum flux occurs within the last 2 years of the NEOWISE survey. This duration cut minimizes significantly the SNe contamination in our sample, since SNe tend to last $\lesssim 6$ months in the optical \citep{2020ApJ...904...35P}, and most become undetectable in the mid-IR after three years  \citep{2016ApJ...833..231T}, unlike IR selected TDEs \citep{masterson_new_2024,2021ApJ...911...31J}.

No specific power-law shape was required for the flux increase or decrease of the transient. This analysis yields a total of 485 TDE candidates.

\begin{figure}

    \includegraphics[width=0.50\textwidth]{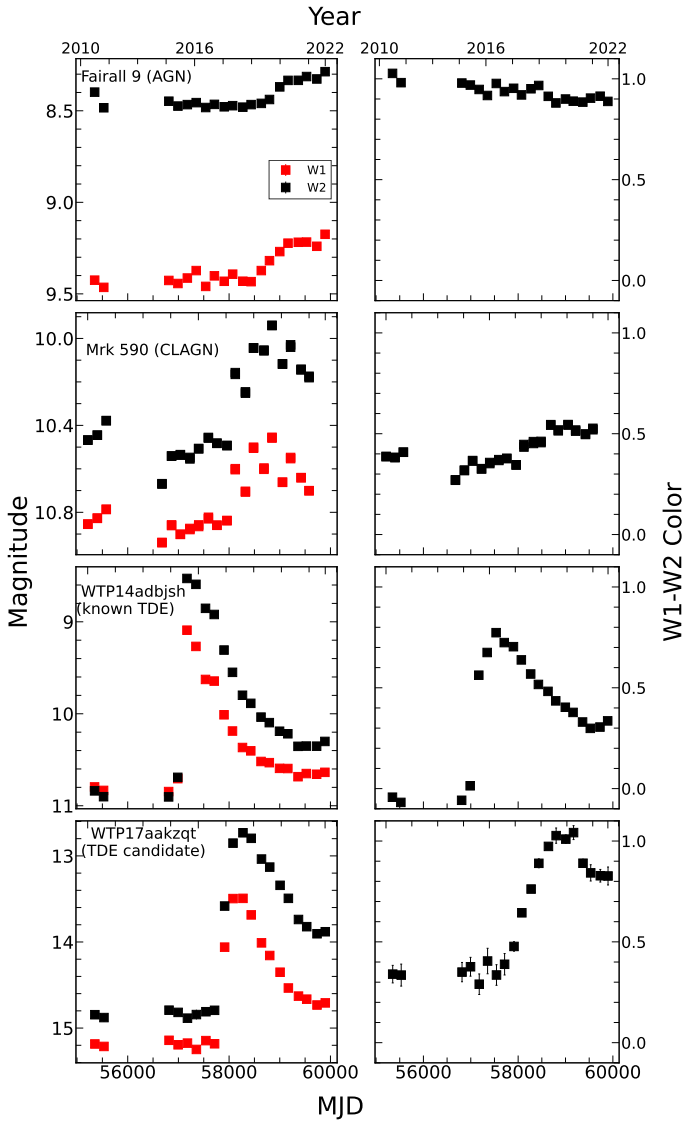}
     
    \caption{Color evolution plots and aperture photometry light curves for Fairall 9 (confirmed AGN), Mrk 590 (confirmed CLAGN), WTP14adbjsh (confirmed TDE), and WTP17aakzqt (categorized as a TDE in our sample). Unlike AGN, TDEs show a clear rise and decline of their W1-W2 colors, due to the accretion event temporarily dominating the observed light.}

\label{fig:color}
\end{figure}

\begin{table*}
  \caption{Selection Process for TDE candidates.}
  \label{tab:selection}
  \centering
  \resizebox{0.7\textwidth}{!}{
  \begin{tblr}{
      colspec={llc},
      row{1}={font=\bfseries}
      }
      \hline
      \hline
      Step&Selection Criteria & No. of objects remaining\\
   \hline
 1&Initial Selection Algorithm&485\\
 2& Magnitude Cut &60\\
 3& Pre-flare W1--W2 Color Cut &37\\
 4& Aperture Flux and Color Evolution &13\\
 5& Cut based on Optical Spectrum &10\\
 
 \hline
\end{tblr}
}
\end{table*}

Then, we chose to focus on the sources with a peak absolute Vega magnitude M$<$-24 in the WISE W2 waveband, as we wish to study highly luminous TDEs. This magnitude corresponds roughly to a lower luminosity bound of $L_\mathrm{peak\, W2} \gtrsim 8\times 10^{43}$ erg s$^{-1}$ for the objects in our subsample. Since the absolute magnitudes were calculated using photometric redshifts, there is an uncertainty associated with this cut-off (\cite{2016MNRAS.460.1371B} estimate uncertainty in photometric redshift to be around $\approx 0.04$, which would lead to an uncertainty in luminosity ranging from a factor of $0.8$ $L_\mathrm{peak\, W2}$ to $0.26$$L_\mathrm{peak\, W2}$ correction for sources with redshift $0.1<z<0.3$; however, we expect that our sources will have a lower bound of about  $L_\mathrm{peak\, W2}\geq 2 \times 10^{43}$ erg s$^{-1}$ (corresponding to the lower limit of the luminosity uncertainty factor derived from the photometric redshift) at least. Applying this peak luminosity threshold also further minimizes the SNe contamination in our sample, as they are typically found to exhibit lower peak luminosities \citep{sukhbold_most_2016}, especially in the mid-IR wavebands \citep{sun_mid-infrared_2022, tinyanont_systematic_2016, 2024ApJ...976..230M}. This smaller sample comprised the brightest 60 sources in the catalog.

Next, we explored the pre-flare $W1-W2$ color for the remaining sources. We calculated the $W1-W2$ color for each object in the first epoch of observation using photometry from the AllWISE program \citep{cutri_explanatory_2013}. Following \cite{stern_mid-infrared_2012}, we decided to exclude transients with a pre-flare color of $W1-W2 > 0.8$, taking this as a clear indicator of AGN activity. 
Accordingly, 23 such candidates were excluded from our sample.

Finally, we used the WISE color evolution of our sources to further distinguish between TDEs and flaring AGN. 
By comparing the color evolution of confirmed mid-IR TDEs \citep{panagiotou_luminous_2023,masterson_new_2024} to those of known AGN, we found that the TDEs feature a characteristic and distinct color evolution, which is naturally explained by their nature of transient accretion systems on top of a galaxy without luminous accretion. In particular, TDEs showed no variation in $W1-W2$ color before the start of the flare, followed by significant increase in $W1-W2$ color (i.e., reddening) during the beginning of the flare. Then, as the transient faded, the $W1-W2$ color decreased. This can be understood as temporary accretion power turning on and nuclear dust echo emission dominating the total mid-IR light from these previously dormant galaxies \citep{Masterson2025ApJ}. This behavior is also in line with the findings of \cite{clark_long-term_2024}, who observe that the $W1-W2$ color of the sources that do not display AGN-like behavior in their sample of extreme coronal line galaxies (ECLEs) systematically decreases with time and falls well below $W1-W2 = 0.8$ over time. \cite{yao_distinguishing_2025} also find that at the rising phase of the flare, mid-infrared color evolution is a suitable metric to distinguish TDEs and other nuclear transients from AGN flares; however, our study also examines color evolution during later portions of the light curve, in order to assess the long-term behavior of the flare and confirm TDEs in a novel way.

Figure \ref{fig:color} shows the NEOWISE $W1-W2$ color evolution plots and W1 and W2 light curves for a typical nearby AGN, Fairall 9 \citep{1977MNRAS.180..391F, 1978Natur.271..334R}, a well studied changing-look AGN (CLAGN), Mrk 590 \citep{2014ApJ...796..134D}, a previously reported mid-IR TDE, WTP14adbjsh in NGC 7392 \citep{panagiotou_luminous_2023}, and a new TDE candidate from our sample, WTP17aakzqt. Fairall 9 shows high $W1-W2$ color, which interestingly decreases as the flux increases. The flare of Mrk 590 around 2020 broadly resembles the light curve shape of a TDE and could thus be mistaken for one if no other information was available.  Its color evolution though is quite distinct from what we expect for a TDE: the pre-flare variations in $W1-W2$ color are comparable in amplitude to the color variation during the flare. WTP14adbjsh, a confirmed TDE, shows almost no variation in color prior to the flare and then shows a rise and fall in color over the duration of the flare. A potential TDE candidate from our sample, WTP17aakzqt, does not display significant color variation prior to the start of the flare, and the source increases in color  by any amount during the flare and then decreases in color as the flare decreases in brightness.
 
Based on these observations and connection to dust echo theory, we removed transients that showed significant variability in either flux or color before the flare. We also required that sources increased in color at the beginning of the flare, followed by a decline in color as the transient faded. Together these cuts brought our sample to a total of 13 TDE candidates.

Three of the objects remaining in our sample, WTP15abycom, WTP17aajhjm, and WTP18aajxru display inconclusive color evolution. In the case of WTP15abycom, its post-flare $W1-W2$ color was lower than its pre-flare $W1-W2$ color, which is characteristic of variability caused by AGN. WTP17aajhjm and WTP18aajxru showed slight variation in magnitude and color prior to the start of the flare. However, we conservatively decided to keep these in our list of objects for spectroscopic follow-up. Through the analysis detailed in Section \ref{sec:spec_analysis}, we were able to confirm that these three sources showed spectral traits characteristic of AGN, which further strengthens the color-based selection criterion we used. The color evolution diagrams for these three sources can be found in Appendix \ref{sec:appendix_ced}.

\subsection{Spectral Analysis Of TDE Candidates} \label{sec:spec_analysis}

Spectra for each of the remaining 13 objects in our sample were collected using the Low Dispersion Survey Spectrograph (LDSS-3) on the Magellan/Clay telescope, the Inamori-Magellan Areal Camera and Spectrograph (IMACS) on the Magellan/Baade telescope, and the Ohio State Multi-Object Spectrograph (OSMOS) on the MDM Hiltner telescope. The LDSS-3 and MDM/OSMOS data were reduced and flux calibrated using a nearby spectroscopic standard star observation taken on the same night with the \texttt{pypeit} \citep{2020JOSS....5.2308P}  python package. The IMACS data were reduced using \texttt{pyraf} \citep{2012ascl.soft07011S} and standard practices for long-slit spectroscopic reduction and flux calibration. Further information on the spectroscopic follow-ups of our TDE candidates, including plots of the spectra, can be found in Appendix \ref{sec:appendix_spectra}.

To fit for the emission line properties, we first  removed the stellar continuum from each spectrum using the Penalized PiXel-Fitting (pPXF) method \citep{2004PASP..116..138C, 2017MNRAS.466..798C, 2023MNRAS.526.3273C}. Key emission lines and telluric regions were masked. The resulting fits returned a normalized emission spectrum for each object, along with noise estimates \citep{2017MNRAS.466..798C}. We then fit Gaussian functions around the peak emission lines \textemdash H$\alpha$ +[N II], H$\beta$, [O III]$\lambda$ 5007, [S II]$\lambda$$\lambda$ 6716,31 and [O I]$\lambda$ 6300\textemdash to estimate the total line flux and line width of each peak. The standard error for the line flux of each peak was calculated by using the estimated standard deviation from the fitted pPXF spectrum to generate a randomized noise distribution for the spectrum. We then resampled each data point within the expected noise, refit the new spectrum, and repeated 1000 such trials to obtain uncertainty measurements on the line flux. The standard deviation in the refitted line flux over all the trials was then taken as the error measurement for the flux.

When fitting the spectra for our candidates, we also sought to eliminate those that had broad spectral features. While TDE spectra can display broad hydrogen lines, they only show them at early times \citep[within 1 year of the flare; e.g.,][]{nicholl_tidal_2019}, whereas all the spectra for our sample were taken 5-10 years after the peak emission. Broad H$\alpha$ and H$\beta$ lines in particular are signs of AGN activity, which we aimed to exclude from our sample. We chose a cut-off line width of $\sigma$ $\geq$ \SI{500}{\km\per\s} to distinguish broad and narrow lines in a given spectrum. 
The objects WTP18aajxru and WTP17aajhjm, which were flagged as possible non-TDEs through photometric analysis in the previous section, were eliminated from our sample due to their spectra displaying such broad lines.

\begin{figure*}
    \centering
    \includegraphics[width=0.48\textwidth]{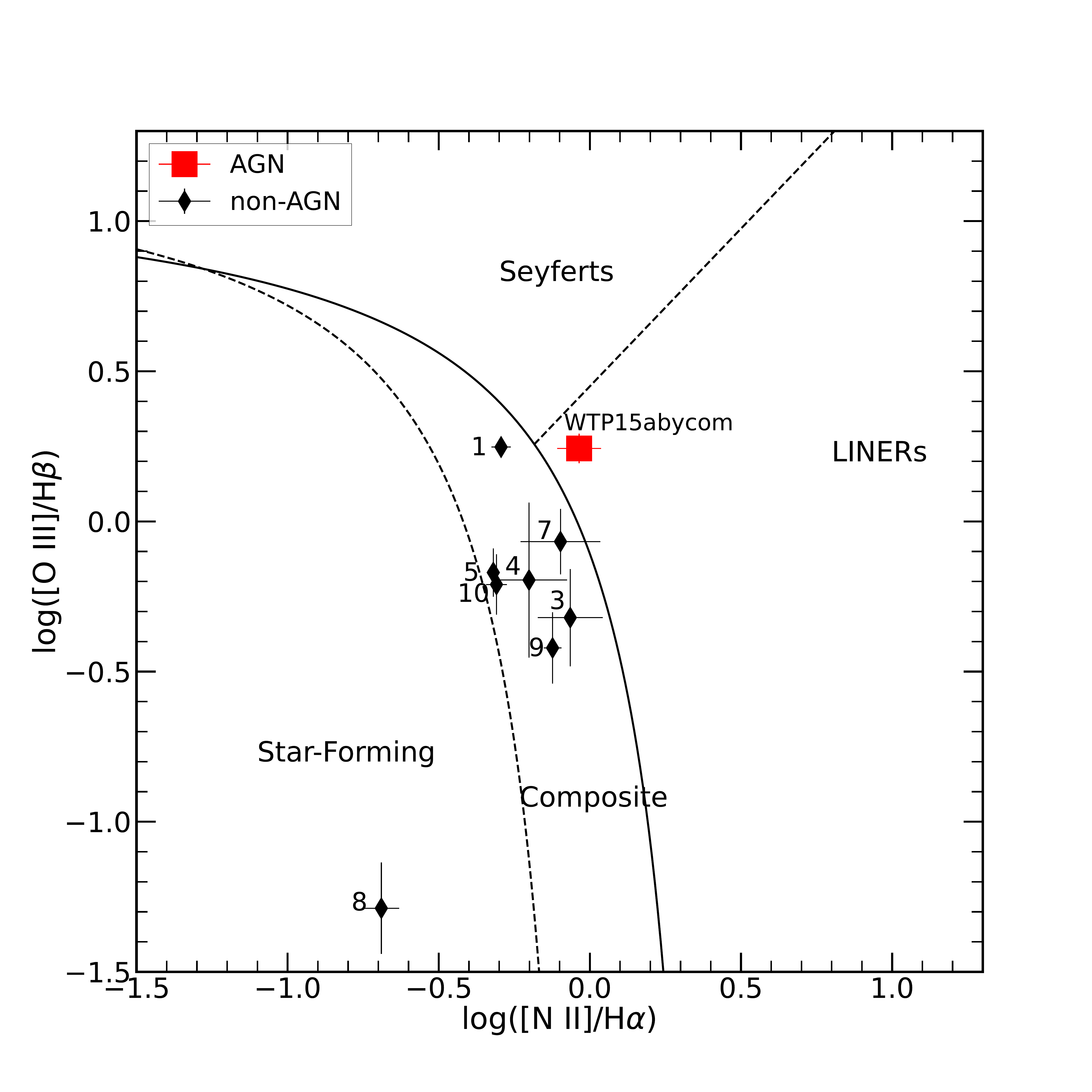}
    \includegraphics[width=0.48\textwidth]{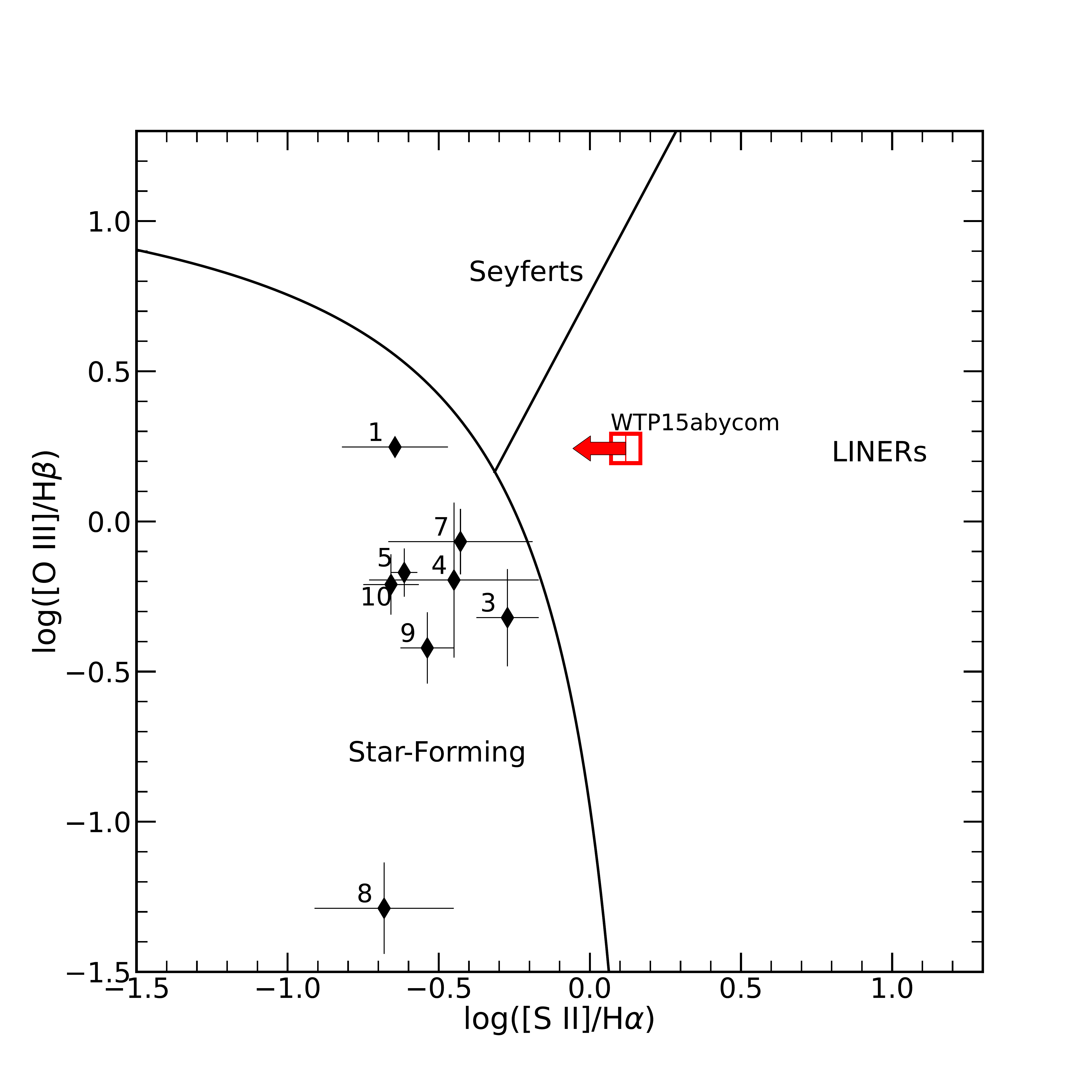}
    \caption{BPT diagnostic plots for the TDE candidates with spectra available. The numbered points correspond to the ID of each object listed in Table \ref{tab:final-TDEs}. The red point labeled as an AGN, WTP15abycom, was previously flagged in our color evolution analysis as a potential non-TDE. In the plot on the right, we use an upper limit to designate WTP15abycom's position as its spectrum did not have a strong [SII] $\lambda$ 6716 emission peak distinguishable from noise.}
    \label{fig:bpt_plots}
\end{figure*}

Using the flux values obtained from the Gaussian fittings, we calculated the flux ratios for the narrow lines to position each object on a BPT diagram \citep{baldwin_classification_1981}. BPT diagrams are frequently used to separate galaxies with AGN from star-forming galaxies, as emission line ratios provide a diagnostic of the underlying ionizing source. 
For instance, objects with a bluer UV spectrum have a higher line ratio of [O III]/H$\beta$, which is characteristic of AGN activity. We created two such diagrams, using the line ratios [O III]/H$\beta$, [N II]/H$\alpha$, and [S II]/H$\alpha$.

The plots were then divided into regions corresponding to the line-ratio boundaries between composite, star-forming, Seyfert and LINER galaxies \citep{kewley_theoretical_2001, kauffmann_host_2003, kewley_host_2006,2007MNRAS.382.1415S}. The final BPT diagrams for the remaining 11 TDE candidates in our sample are displayed in Figure \ref{fig:bpt_plots}. As shown in the figure, the majority of the objects in our final sample fall in the composite/star-forming region of the diagram, which indicates an elimination of AGN contamination in our final sample. The one source labeled as an AGN is WTP15abycom, which we had previously flagged as being a potential non-TDE from the color evolution analysis in Section \ref{sec:photo_analysis} and which we exclude from the subsequent analysis.

It should be noted that the spectra for two of the objects in our sample, WTP14adauec and WTP17aaiyxe, were not of sufficient quality to apply the aforementioned pPXF and BPT analysis. Nevertheless, we were able to determine the spectroscopic redshift of these objects and confirm their lack of broad features. We chose to follow a conservative approach of including them in our TDE sample for future analysis. As will be discussed later in Section \ref{sec:lum_func}, we also test our final results by assuming these are AGN and excluding them from the sample, and our conclusions remain the same qualitatively. 

The above analysis using photometric  and spectroscopic data resulted in a final sample of 10 TDE candidates, which are listed in Table \ref{tab:final-TDEs}. The table includes host galaxy information, their date of first NEOWISE detection, redshift values obtained from spectroscopic observation, as well as information about multi-wavelength counterparts, further detailed in Section \ref{multi-wav}.

\begin{table*}
\caption{Final IR TDE Sample}
  \centering
  \resizebox{1\textwidth}{!}{
  \begin{tblr}{
      colspec={lcrrccc},
      row{1}={font=\bfseries},
      row{2}={font=\bfseries}
      }
    \hline
    \hline
    
            ID& WTP Name & RA(deg.)  & Dec(deg.)  & Host& Date of First & Optical \\
            &&&&Galaxy&WISE Detection&Counterpart \\

    \hline
    1&WTP14actvjz& 56.086780 & -49.8646&2MASS J03442084-4951527&2015-03-14&No\\
    2&WTP14adauec&  145.1229851 & 37.135161&WISEA J094029.53+370806.4&2014-11-11 &No\\
    
    3&WTP15abwlqq& 307.784893& -50.6332941&WISEA J203108.40-503759.8&2015-04-17&No\\ 
    
    4&WTP15abzxqe& 228.4378545& -86.740893473&WISEA J151344.79-864425.5&2015-09-08&No\\
    5&WTP16aapkfo&  48.966392 & -0.9921&WISEA J031551.85-005931.7&2016-01-25&No\\
    
    6&WTP17aaiyxe& 229.341644 & 46.9701932&WISEA J151722.00+465812.6&2017-01-14&Yes\\
    7&WTP17aakzqt&  45.697713 & -65.742939 &WISEA J030247.52-654433.6&2017-06-08&No\\
    8&WTP17aanbho& 346.6021038& -7.272606812&WISEA J230624.44-071621.2&2017-11-18&Yes (AT2017fot)\\
    9&WTP18aamykg&  53.707291 & 25.033717 &WISEA J033449.79+250201.1&2018-08-20&No\\
    10&WTP19aajfdd& 60.46709897& -24.4161968&WISEA J040152.06-242458.0&2019-01-21&No\\
    
    \hline
    
\end{tblr}
}

\label{tab:final-TDEs}
\end{table*}

\section{Searching for Multi-Wavelength Counterparts} 
\label{multi-wav}

Since the mid-IR emission in TDEs arises from reprocessed X-ray, optical and UV emission, we searched for multi-wavelength counterparts of each object in Table \ref{tab:final-TDEs} to gain a more robust understanding of these events. Primarily, X-ray and optical surveys were used to search for counterparts in our sample, as detailed in the subsections below. We also attempted to search for radio counterparts to our flares, as it has been found that a significant fraction of TDEs display some radio emission, with some even producing relativistic jets  \citep{alexander_radio_2020}.
We searched for radio observations in the Very Large Array Sky Survey \citep{lacy_karl_2020}, using the Canadian Initiative for Radio Astronomy Data Analysis (CIRADA) Cutouts service\citep{gordon_catalog_2020}, but found no detections for any of our events in the survey data.

\subsection{X-Ray Counterparts} \label{x-ray}

To search for X-ray transients, we primarily used the publicly released data from eROSITA survey \citep{predehl_erosita_2021}, eROSITA-DE DR1, which includes survey data collected from December 2019 to June 2020. Out of the 10 final candidates in our sample, 7 were located in the eROSITA-DE sky region. None of these objects had a positive eROSITA detection within 15 arcseconds of their given position. Using the eROSITA DR1 upper limits search tool\footnote{https://erosita.mpe.mpg.de/dr1/erodat/upperlimit/single/}, we find that our sources have upper limits between $0.63-2\times10^{-13}$ $\mathrm{erg}$ $\mathrm{cm^{-2}}$ $\mathrm{s^{-1}}$ in the $0.2-5$ keV band. We note that the DR1 survey took place at least 2 years after most of the flares in our final sample, so the lack of eROSITA detections does not conclusively indicate a lack of X-ray activity for these TDEs. In addition, the distance of these sources may result in their late time X-ray emission being fainter than the eROSITA detection threshold. This may also explain why a larger fraction of nearby mid-IR selected TDEs was found to be X-ray bright a few years after their peak \citep{masterson_new_2024}. 

We also looked for ROSAT survey \citep{voges_rosat_1999} detections for these candidates. Since the ROSAT all-sky survey concluded in 1991, decades before WISE observations of our sources began, any positive detections in ROSAT would eliminate those potential candidates from being TDEs uncontaminated by AGN activity. No detections for our sample were found in the ROSAT survey.

In addition, we searched for X-ray detections from XMM-Newton \citep{jansen_xmm-newton_2001}, Swift \citep{burrows_swift_2005}, \citep{weisskopf_chandra_2012} and NuSTAR \citep{harrison_nuclear_2013} using NASA's High Energy Astrophysics Science Archive Research Center (HEASARC) search tool\footnote{https://heasarc.gsfc.nasa.gov/cgi-bin/W3Browse/w3browse.pl}. Of the sources in our sample, only one was serendipitously in the field of view of one past XMM-Newton observation in 2021. We followed the standardized analysis and deduced that no X-ray bright source was associated with the position of WTP16aapkfo.
The other sources in our sample have not been previously observed by current X-ray observatories.

\subsection{Optical Counterparts} 
\label{optical}

We searched the Transient Name Server (TNS)\footnote{https://www.wis-tns.org/} for previously reported optical transients corresponding to our WISE objects. Of the candidates in our sample, only WTP17aanbho has a reported optical counterpart on TNS, AT2017fot, as shown in Table \ref{tab:final-TDEs}. AT2017fot, internal name PTSS-17tts, was discovered by the Schmidt telescope at Xuyi Observatory on 2017-07-20 as part of the PMO-Tsinghua Transient Survey(PTSS), with a Vega magnitude of 19.04 in the Sloan-i filter 
\citep{xu_ptss_2017}. 

\begin{figure*}

    \includegraphics[width=0.52\textwidth]{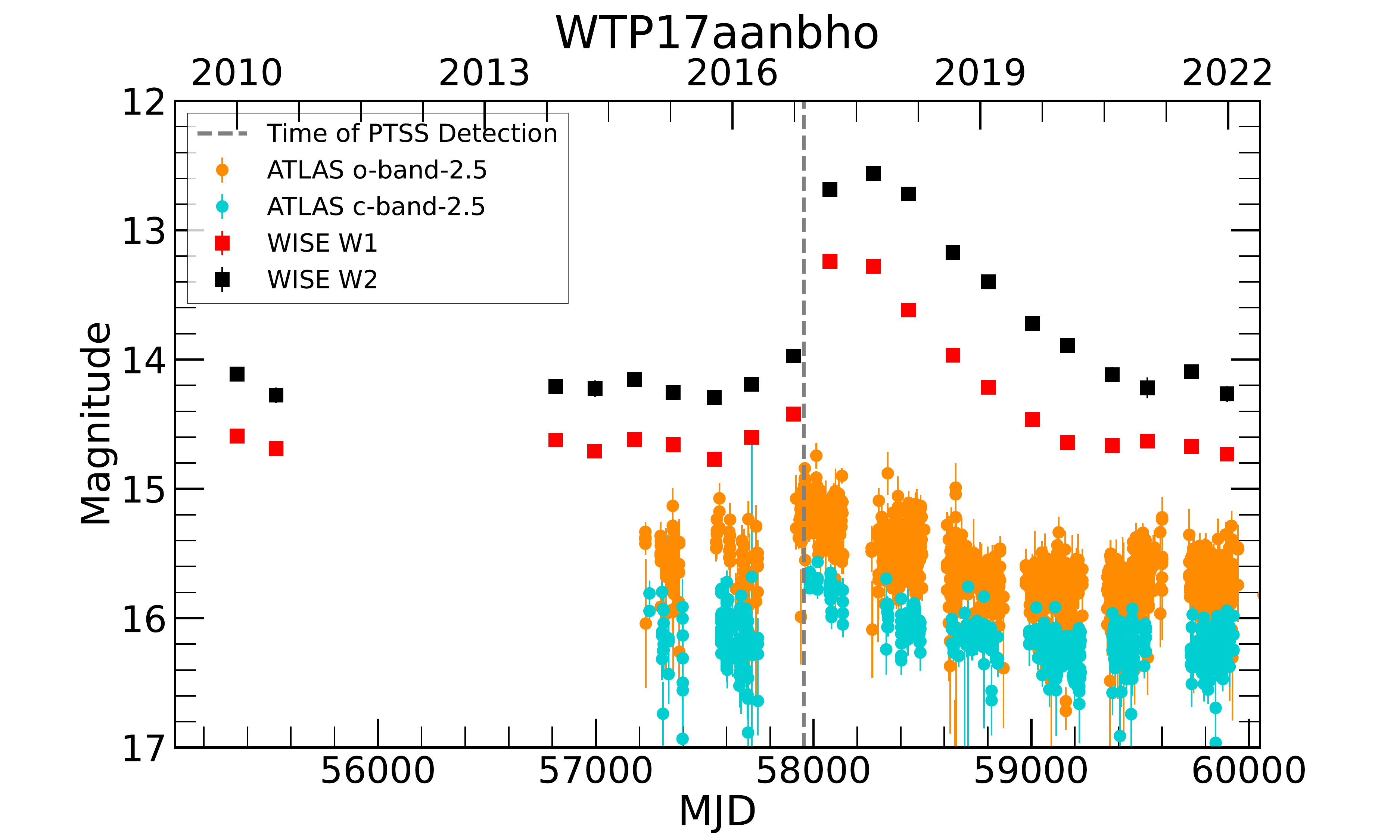}
    \includegraphics[width=0.52\textwidth]{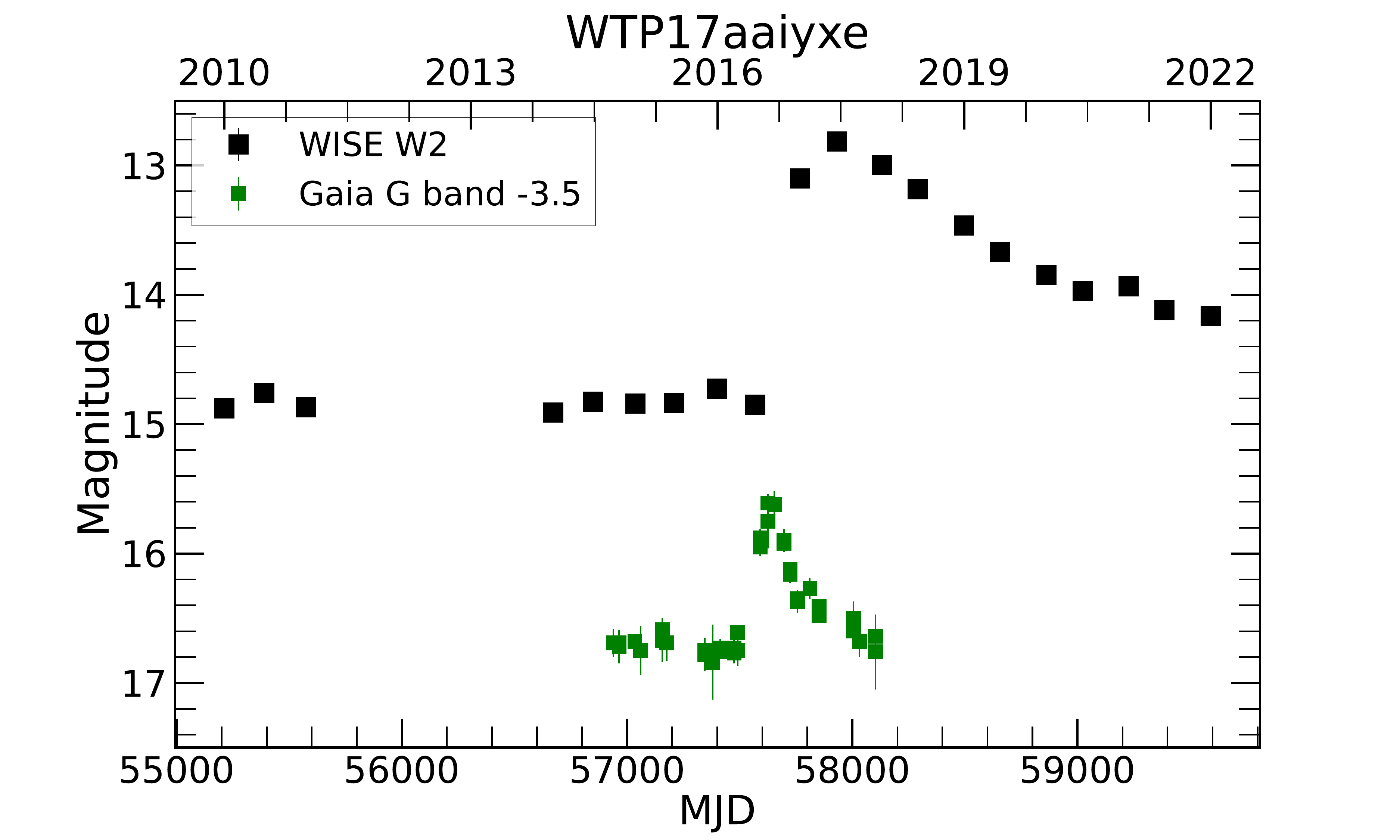}
    \caption{Light curves of WTP17aanbho (ATLAS forced photometry and WISE aperture photometry) and WTP17aaiyxe (GAIA G and WISE W2 aperture photometry). The ATLAS data was plotted with a magnitude offset of 2.5, and the GAIA light curve was plotted with a magnitude offset of 3.5. The time of the  PTSS detection of AT2017fot (the optical counterpart of WTP17aanbho reported on TNS) is denoted by the dark grey dashed line in the first plot. Both WISE flares are notably brighter than their optical counterparts.} 
    \label{fig:optical}
\end{figure*}

In addition, the ASAS-SN \citep{shappee_man_2014}, ATLAS \citep{tonry_atlas_2018} and ZTF \citep{bellm_zwicky_2019} surveys were used to further search for optical variability in our TDE candidates. 
Using the ATLAS forced photometry pipeline, we found that WTP17aanbho also displayed a TDE-like optical light curve in the ATLAS o-band and c-band filters. Figure \ref{fig:optical} shows the WISE aperture photometry light curves for WTP17aanbho along with the ATLAS AB magnitude light curves (which are plotted with a magnitude offset of 2.5). The optical activity begins slightly before the infrared emission takes place, which follows our understanding that IR emission is sourced from reprocessed optical/X-ray emission. We also note that the optical flare is considerably dimmer than the WISE flare, which is expected for obscured sources such as these. 
Since ASAS-SN only probes down to approximately 17 mag in the Johnson V filter \citep{kochanek_all-sky_2017}, most of our sources were too dim to obtain a clear magnitude light curve using its aperture photometry pipeline. We also searched the ZTF survey for potential optical counterparts to our transients. It should be noted that ZTF came online in 2018, well after most of the flares in our sample took place. Nevertheless, none of our sources in the ZTF sky footprint displayed significant optical variability.

Finally, we found that WTP17aaiyxe has a corresponding optical transient detected by the Gaia mission \citep{gaia_collaboration_gaia_2016}. 
\cite{kostrzewa-rutkowska_gaia_2018} reported this object as a nuclear transient, labeling it \text{GNTJ151722.00+465812.67}, which could either be a supernova or a TDE, without any attempts at further classification. The Gaia light curve (with a magnitude offset of 3.5) from \cite{kostrzewa-rutkowska_gaia_2018} along with the WISE W2 aperture photometry light curve is shown in Figure \ref{fig:optical}. Again, the optical transient slightly precedes the IR flare.

\section{Luminosity Function and Rate Calculation} \label{sec:lum_func}

\begin{table*}
  \centering
  \caption{$D_{\max}$, limiting luminosity and peak luminosity values for the sample.}
  \resizebox{0.8\textwidth}{!}{
  \begin{tblr}{
      colspec={lccc},
      row{1}={font=\bfseries},
      row{even}={bg=white!10}}
    \hline
    \hline
            \text{WTP Name} & \text{redshift} ($z$) & $D_{L}$ (Mpc) & \textbf{$L_\mathrm{peak\, W2}$}\text{(erg s$^{-1}$)}& \textbf{$L_\mathrm{lim\,W2}$} \text{(erg s$^{-1}$)}& $D_{\max}$ (Mpc) \\
    \hline
    WTP14actvjz&0.225 & 1155.1 &$7.03\times10^{43}$ &$4.51\times10^{43}$& $3749.0$ \\
    WTP14adauec&0.120& 578.5 &$3.82\times10^{43}$&$3.39\times10^{43}$ &$2708.4$ \\ 
    WTP15abwlqq	&0.180& 900.5 &$6.21\times10^{43}$& $4.53\times10^{43}$&$3369.6$\\ 
    WTP15abzxqe	& 0.186& 933.8 &$6.93\times10^{43}$& $5.66\times10^{43}$&$3980.0$\\ 
    WTP16aapkfo	& 0.279& 1474.5  &$1.03\times10^{44}$&$5.60\times10^{43}$&$3073.9$\\ 
    WTP17aaiyxe	& 0.154& 758.4 &$3.79\times10^{43}$& $2.41\times10^{43}$&$2456.5$\\ 
    WTP17aakzqt	& 0.210& 1069.0 &$1.46\times10^{44}$&$1.05\times10^{44}$&$5389.2$\\ 
    WTP17aanbho	& 0.111&532.0&$3.02\times10^{43}$&$1.38\times10^{43}$&$1378.7$\\ 
    WTP18aamykg	& 0.202&1023.6 &$6.6\times10^{43}$&$5.73\times10^{43}$&$2837.2$\\ 
    WTP19aajfdd	& 0.175&872.9&$3.76\times10^{43}$&$2.85\times10^{43}$&$2188.8$\\ 
    \hline
\end{tblr}    
}

\label{tab:lum_dist}
\end{table*}

To estimate the rate of highly luminous mid-IR TDEs, we followed the classical ``1/$V_{\max}$" approach of \cite{schmidt_space_1968}. This approach for estimating the luminosity function of TDEs has been presented in detail in a number of recent papers \citep[see for example][]{yao_tidal_2023}. 

The rate $\phi_{j}$ for a given luminosity bin $j$ with width $\Delta_{j}\log L_\mathrm{peak\, W2}$ containing $n_{j}$ TDEs was calculated using 
\begin{equation}
\phi_{j}=\frac{\sum_{i=1}^{n_{j}} \frac{1}{(T_{\mathrm{span}}V_{\max,i})}}{\Delta_{j}\log L_{\rm peak\, W2}},
\label{eq:phi}
\end{equation}
where $T_{\mathrm{span}}$ is the duration of the experiment or survey and $V_{\max,i}$ is the volume of the maximum distance out to which a transient could be detected. Our selection criteria largely flagged events that have positive detections between 2014 and 2021; hence, we set $T_{\mathrm{span}}$ to 7 years for this rate calculation. The volume $V_{\max,i}$ is defined by 
\begin{equation} V_{\max,i}=\frac{4\pi}{3}D_{\max,i}^3,
\end{equation}

where $D_{\max,i}$ refers to the maximum distance out to which a transient object can be detected above the flux limits of WISE. $D_{\max,i}$ was computed by estimating a flux limit for each object and then computing the maximum distance at which the transient could be detected given that limit. 

In our detection pipeline, we require that a TDE candidate is detected in difference photometry at $\geq 5\sigma$ for 4 consecutive observations, i.e. 2 years. Therefore, we use the fourth largest luminosity value to calculate $D_{\max}$, rather than simply the peak luminosity. We define this luminosity as the limiting luminosity $L_\mathrm{lim\,W2}$. $D_{\max,i}$ was then calculated using this limiting luminosity and the survey's flux limit, which was typically around 0.08 mJy for a $5\sigma$ detection. Given our small sample size, we opt to divide the 10 TDE candidates into two bins, sorted by log luminosity ($L_\mathrm{peak\, W2}$) such that each bin has five sources. Finally, we estimate uncertainties on $\phi_j$ using Poisson error statistics \citep{gehrels_confidence_1986}.
Table \ref{tab:lum_dist} compiles the limiting luminosity and $D_{max}$ values for the objects in our sample.

It is important to note that the above calculations do not factor in the completeness of the survey. We refer to the recent GLADE+ galaxy catalog companion work \citep{dalya_glade_2022} to estimate the sample completeness.
In our final sample, the closest source is at around $532$ Mpc, while the farthest source is at over $1400$ Mpc. Figure 2 of \cite{dalya_glade_2022} shows that the completeness at these distances, in terms of the average luminosity of the GLADE+ galaxies, ranges from around $40$\% to below $20$\%. However, our sample consists of bright galaxies, most likely because we have selected the most luminous events, and so we expect a higher completeness in our case. Additionally, the GLADE+ completeness extends only to $800$ Mpc, whereas seven of our sources lie beyond that.

Although there are no exact measures for the completeness of our sources with respect to their luminosity and distance, we attempt to account for completeness changing as a function of luminosity distance by approximating an exponential decay for the GLADE+ completeness between 400 and 800 Mpc. Therefore the new rate taking into account this completeness correction becomes:
\begin{equation}
\phi_{j}=\frac{\sum_{i=1}^{n_{j}} \frac{1}{(T_{\mathrm{span}}V_{\max,i})}\times CF(D_{L, i})}{\Delta_{j}\log L_{\rm peak\, W2}},
\label{eq:phi}
\end{equation}
where $CF(D_{L, i})$ represents the completeness correction factor at the luminosity distance of a given source.
This correction results in a final overall IR TDE rate of $1.2^{+0.5}_{-0.4}\times10^{-10}$ Mpc$^{-3}$year$^{-1}$ in the luminosity range $L_\mathrm{peak\, W2} = 3 \times 10^{43} - 1.5 \times 10^{44}$ erg s$^{-1}$.

As mentioned in Section \ref{sec:spec_analysis}, we did not complete the BPT analysis for two objects in the final sample, WTP14adauec and WTP17aaiyxe. To test our results, we also calculated the TDE rate by excluding these objects from our sample. The corrected TDE rate for this subset of sources is $1.05^{+0.5}_{-0.4}\times10^{-10}$ Mpc$^{-3}$year$^{-1}$.
 We therefore conclude that including these objects in our final sample does not qualitatively change our results.

\begin{figure*}
\centering
\includegraphics[width=0.9\textwidth]{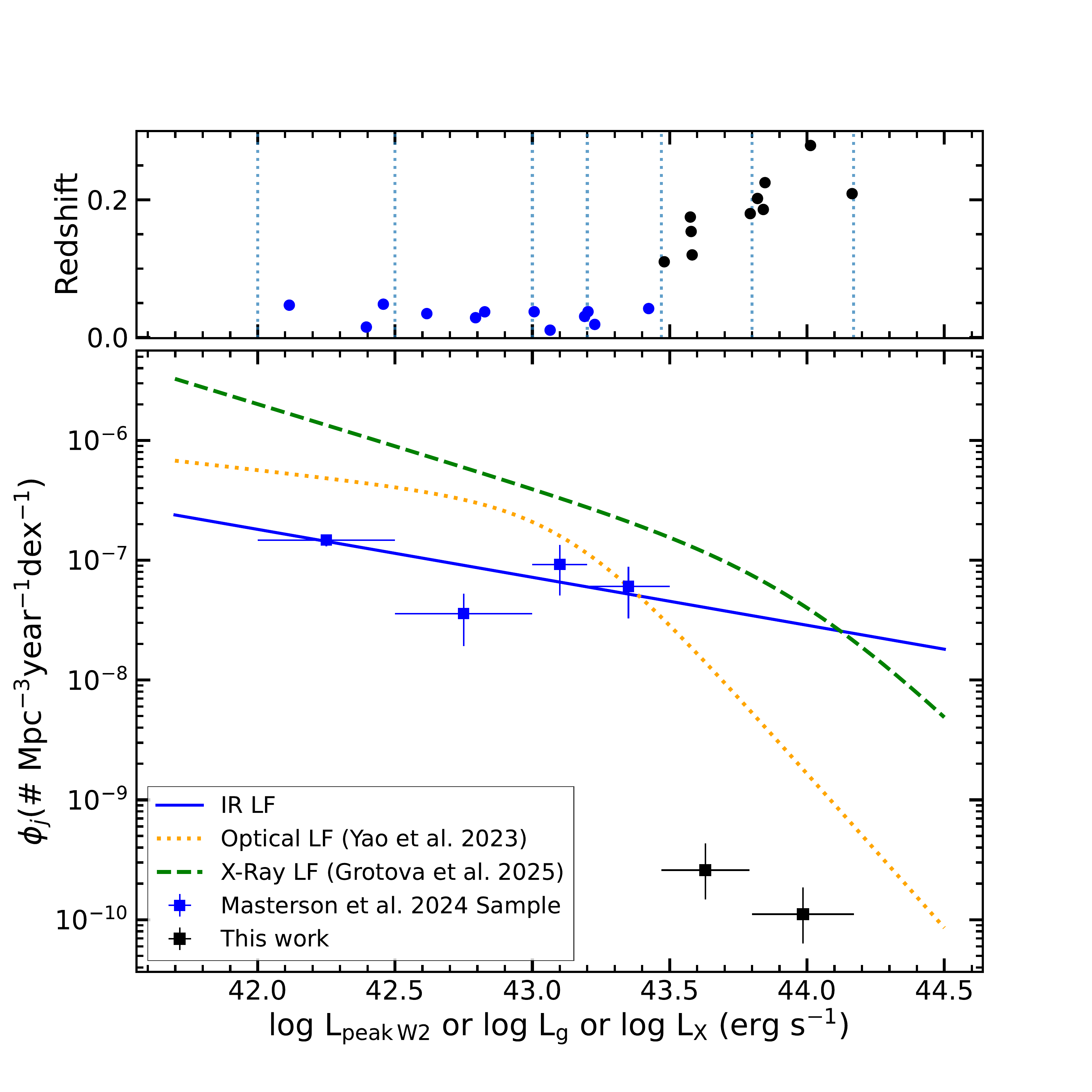}
    \caption{Luminosity function of mid-IR TDEs. The top panel shows the redshift of our sources versus log $L_\mathrm{peak\, W2}$, with the vertical lines representing the boundaries of the luminosity bin each source was placed in below. The lower panel shows the calculated IR TDE rate per bin versus log $L_\mathrm{peak\, W2}$ for our sample in black. We have also plotted the X-ray luminosity function from \cite{grotova_population_2025} in green, the optical luminosity function from \cite{yao_tidal_2023} in orange, and the mid-IR luminosity function from \cite{masterson_new_2024} in blue. A clear break in the mid-IR TDE luminosity function is observed.} 
\label{fig:lum_func}
\end{figure*}

The lower panel of Figure \ref{fig:lum_func} displays the TDE rate $\phi_j$ versus $\log L_\mathrm{peak\, W2}$ for two luminosity bins in our sample, as well as the luminosity function estimated by \cite{masterson_new_2024} for less luminous mid-IR-selected TDEs. It is evident that the TDE rate decreases significantly at higher luminosities, with a turn-over around $\log (L_\mathrm{peak\, W2} / \mathrm{erg\, s}^{-1}) \simeq 43.4$. To compare our luminosity function to past work, we extended the single power-law model of \cite{masterson_new_2024} for the mid-IR luminosity function to the higher luminosities probed by this work. The event rate derived from our sample is at least a factor of $2 \times10^2$ lower than what is predicted by the single power-law model. We find it almost impossible that this feature is a result of any uncertainty with regard to sample completeness in our calculations, as a much higher correction to the rate would still result in the detection of such a break. 
Our qualitative results remain unaffected by the completeness corrections applied. Further, the turn-over luminosity is lower by almost an order of magnitude from the break seen in the luminosity function of galaxies \citep[e.g.][]{Dai2009,Lake2018}, and is thus most likely inherent to the TDE population.

\section{Discussion}
\label{sec:disc}

\subsection{Interpreting the Turn-Over in the Luminosity Function}
\label{sec:disc_cutoff}

A natural explanation of the observed turn-over in IR TDE luminosity function stems from assuming that TDEs are Eddington limited and as a result more luminous sources correspond on average to systems with larger black hole masses. This behavior is in line with observational analyses from past surveys: in their study of non-Seyfert galaxies that display slow-declining MIR/WISE light curves, \cite{2018MNRAS.477.2943W} conclude that the peak IR luminosity of the IR flare correlates with black hole mass. Then, it becomes straightforward to connect the observed turn-over to the suppression of TDEs in larger black hole masses \citep[e.g.][]{Velzen2018}, as above a mass limit ($\approx 10^8 M_\odot$ for non-spinning black holes), the tidal radius is smaller than the event horizon and no electromagnetic emission is expected. 

To further test this, we estimated the SMBH mass of our sample using the scaling relation between the black hole mass and the stellar mass of the host galaxy \citep{2020ARA&A..58..257G}, with the latter obtained from the GLADE+ catalog. We calculated a mean black hole mass of log $\mathrm{M_{BH}\approx8.01\pm0.13 \, M_\odot}$ for our sources, fully consistent with the aforementioned limit. In comparison, we calculated a mean mass of log $\mathrm{M_{BH}\approx7.36\pm0.12 \,M_\odot}$ for the gold IR TDE sample in \cite{masterson_new_2024}, which further corroborates the thesis that we here probe a sample of more massive black holes.

A similar turn-over or break has been observed in the luminosity function of optical \citep{Velzen2018, yao_tidal_2023} and X-ray \citep{Guolo2024,grotova_population_2025} bright TDEs, two examples of which are plotted in Fig. \ref{fig:lum_func} as well. These breaks have also been associated with the suppression of TDEs around more massive SMBHs.

It is interesting to note that the exact value of the turn-over varies for different wavebands. For instance, \cite{yao_tidal_2023} detect a break in the optical luminosity function from their sample at $\log (L_{g} / \mathrm{erg\, s}^{-1}) \simeq 43.1$ and \cite{Guolo2024} measure a break of $\log (L_{X} / \mathrm{erg\, s}^{-1}) \simeq 44.1$ for their X-ray luminosity function. The break in our luminosity function is intermediate between these two values, falling around $\log (L_{\rm peak\, W2} / \mathrm{erg\, s}^{-1}) = 43.5$. Interestingly, this is comparable to the break found in the peak W2 luminosity function constructed by the MIRONG survey \citep{jiang_mid-infrared_2021}. The observed turn-over in the MIRONG luminosity function is not as strong as what we obtain for our sample, likely because that sample includes TDE candidates as well as turn-on/changing-look AGN, whereas our study focuses extensively on excluding AGN flares from the final sample. Overall, we suspect that the different turn-over luminosities is a manifestation of different bolometric corrections for the emission mechanisms in different wavebands. Ultimately, constructing an SMBH mass function from a flux-limited, multi-wavelength sample of TDEs will be the ideal way to assess the suppression level at high SMBH masses. 

For our sample of IR-selected TDEs, the theory of dust echoes can help us understand this relatively low cutoff luminosity. For a TDE with black hole mass $10^8 M_\odot$, the Eddington luminosity is $L_{\rm Edd} \sim 10^{46}$ erg s$^{-1}$. Assuming a surrounding dusty medium with covering factor of 0.5, simple dust reprocessing models suggest that the mid-IR emission will then feature a peak luminosity of $L_{\rm peak\, W2} \sim 10^{43.5-44}$ erg s$^{-1}$ in the W2 band \citep{Masterson2025ApJ}, which is comparable to the observed turn-over luminosity for mid-IR TDEs. The difference between the peak W2 and peak bolometric luminosities is a result of dust emission peaking in larger wavelengths than the W2 waveband and of this dust emission originating from much larger areas than the intrinsic emission. In particular, the larger emitting area associated with the dust emission extends the light curve, yielding lower peak luminosities as the emission is spread out over a longer time period. 

With the dust emission in the IR, it is important to consider the effects of sublimation on the observed IR flares and consequent luminosity function. We expect the rise time of the MIR light curve to be set by the physical light travel time to the dust sublimation radius, $t_\mathrm{rise} \sim r_\mathrm{sub}/c$. The sublimation radius for ISM-like dust scales like $r_\mathrm{sub} \propto L_\mathrm{peak,\, bol}^{1/2}$ \citep[see Eq. (1) in ][]{vanVelzen2021}, meaning that more luminous flares should have longer rise times. Surprisingly, the rise times of these flares are not significantly longer than those presented in \cite{masterson_new_2024}. This may suggest a stronger dependence on the parameters of the nuclear environment \citep[e.g.][]{2025ApJ...989...27T} or temperature-dependence, as WISE light curves are not always representative of the total IR flux; we will explore this in future work. Likewise, the larger sublimation radii of luminous flares could dilute their peak IR luminosity. To test whether this can produce an artificial cut-off in the luminosity function, we used the model presented in \cite{Masterson2025ApJ}, which produces time-dependent MIR spectra based on a TDE-like flare propagating through a dusty shell that is optically thin to its own emission. We found no strong cut-off in the IR luminosity function (see Appendix \ref{sec:appendix_dust}), and thus, we do not expect that sublimation effects are responsible for the observed cut-off in the luminosity function.

\subsection{Mid-IR Color Evolution for IR TDE Selection}
\label{sec:disc_color}

In this work, we present a new way for vetting TDE candidates based on the MIR color evolution. In particular, we require the transient's host reddens as the flares onsets before returning to bluer colors during the fainting of the flare. Our methodology for TDE selection is motivated by dust echo physics and effectively eliminates many AGN. The significant reddening during the early portion of the flare stems from the newly-formed accretion flow illuminating hot dust in the vicinity of the SMBH. As the intrinsic optical/UV/X-ray light from the TDE fades, the dust cools and the IR flux fades. This leads to the host galaxy dominating again, and the total IR color returns towards the pre-outburst value. Interestingly, many of the TDEs do not reach their pre-flare colors within the WISE lifetime; this suggests that the transients are still on, consistent with theory for long-lived TDE accretion disks \citep{Mummery2020}, late-time UV plateaus in optically-selected TDEs \citep{vanVelzen2019,Mummery2024}, and the short wavelength excesses seen in the JWST spectra of IR-selected TDEs \citep{Masterson2025ApJ}. 

An important take-away of this work is that we can effectively use mid-IR color evolution to distinguish between TDEs and flaring AGN. Notably, our final selection step, based on the WISE color evolution, yielded an extremely high purity of $10/13 \approx 80\%$. This will reduce the need for follow-up spectroscopy, which is going to be vital for searches for obscured TDEs in next-generation IR surveys \citep[e.g., the Nancy Grace Roman Space Telescope,][]{Spergel2015} that promise to detect fainter and more distant transients.

\section{Summary and Conclusion}
\label{sec:conclusion}

In this work, we aimed to constrain the high-end of the luminosity function of mid-IR-selected TDEs. We performed a systematic analysis of the NEOWISE archive and applied a range of selection criteria to robustly identify TDEs. We present a sample of 10 IR bright TDEs, with peak luminosity of $\log L_\mathrm{peak\, W2}$ $>43.4$.

In our effort to identify TDEs, we have developed a new selection method based on the IR color evolution of the host galaxy, which was able to distinguish well between TDEs and AGN flares. Although past works have used pre-flare and early-time $W1-W2$ color to separate AGN flares \citep{clark_long-term_2024,yao_distinguishing_2025}, this study presents a novel way to distinguish IR TDEs from AGN based on $W1-W2$ color evolution over the entirety of the IR flare. The TDE hosts appear redder as the flare evolves, with their color peaking at a value close to $W1- W2=0.8$, before returning towards their pre-flare bluer color. We envision that this criterion will be useful for selecting TDEs in current and upcoming IR survey missions, such as SPHEREx \citep{2014arXiv1412.4872D}, the Roman Space Telescope \citep{Spergel2015} and NEO Surveyor \citep{Mainzer2023}, minimizing considerably the observational resources required to classify nuclear transient events.

We explored archival data and past literature to search for multi-wavelength counterparts to our transient events. Only two of our sources, WTP17aaiyxe and WTP17aanbho, had reported optical counterparts. In both cases, the optical activity preceded the IR flare, which is in line with  our understanding that IR emission in TDEs is partially derived from reprocessed optical emission.

We derive a total volumetric TDE rate of $1.2(^{+0.5}_{-0.4})\times10^{-10}$ Mpc$^{-3}$year$^{-1}$ for our sample of highly luminous IR TDEs. This rate is several orders of magnitude lower than the volumetric rate estimated for less luminous IR bright TDEs, highlighting a significantly suppressed rate of events above $\log L_\mathrm{peak\, W2} \simeq 43.4$. The detected turn-over adds to similar results found in other wavebands, and confirms that the detected flares are the result of stars being tidally disrupted by a supermassive black hole. This result motivates further investigations of the luminosity function of IR bright TDEs, which will allow us to probe the mass function and potentially spin distribution of supermassive black holes. We plan to present such an analysis for a flux-limited sample of NEOWISE detected events in a future publication.

\begin{acknowledgments}
MM thanks Wenbin Lu for insightful discussions. We acknowledge the support of the National Aeronautics and Space Administration through ADAP grant number 80NSSC24K0663. This work is based on observations obtained at the MDM Observatory, operated by Dartmouth College, Columbia University, Ohio State University, Ohio University, and the University of Michigan.
\end{acknowledgments}

\appendix
\section{WISE Light curves of Final TDE sample}
\label{sec:appendix_lc_plots}

\begin{figure*}

\centering
\includegraphics[height=0.95\textheight]{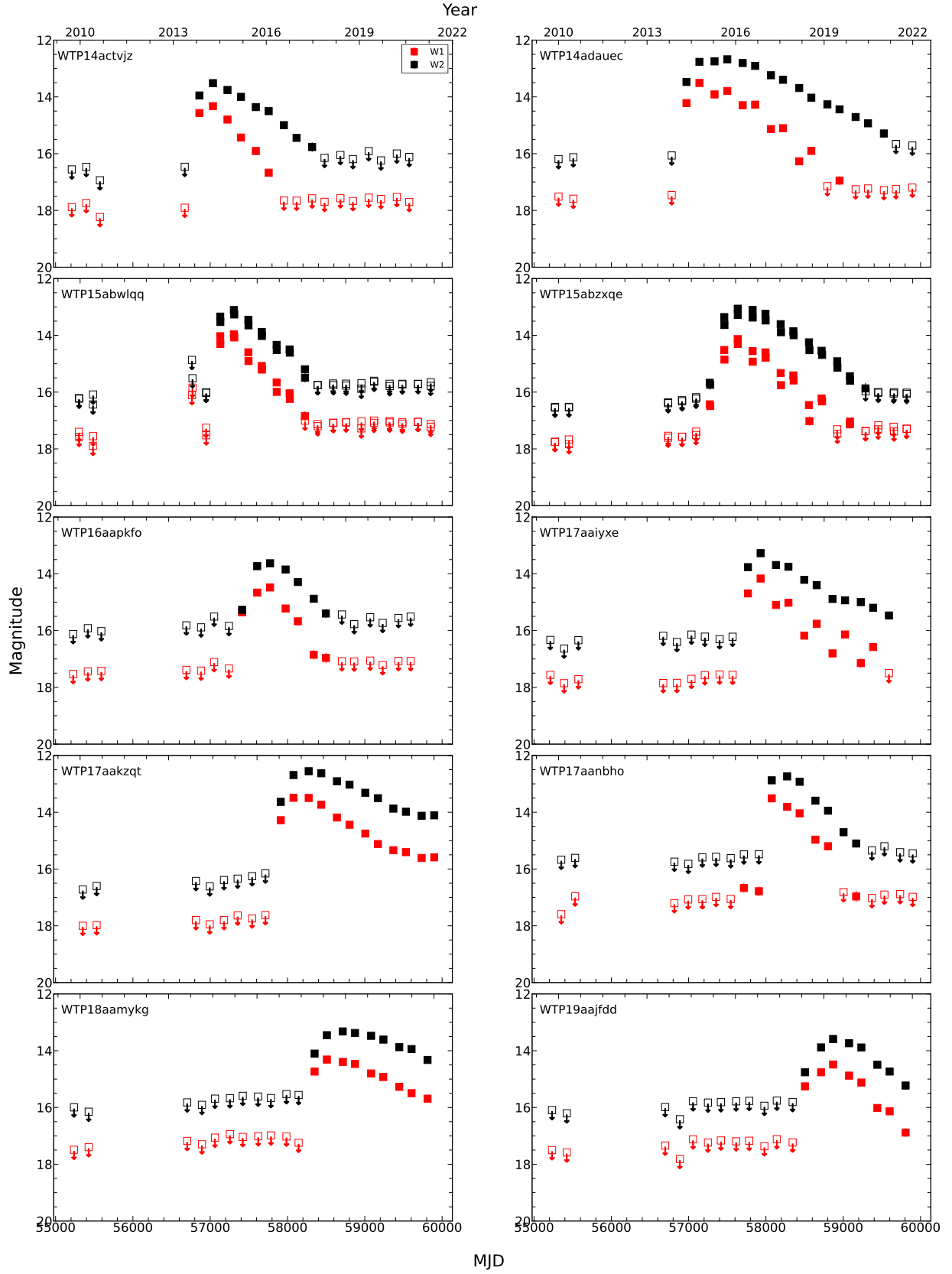}
    \caption{Difference light curves of the final TDE sample, also listed in Table \ref{tab:final-TDEs}.Upper limits are denoted by empty markers with downward arrows and detections are shown as filled squares with error bars.}
\label{fig:all_TDE_lcs}
\end{figure*}

We present the full W1 and W2 magnitude light curves of the objects in our final sample of TDEs in Figure \ref{fig:all_TDE_lcs}. These light curves were plotted using difference photometry obtained from our pipeline.

\section{Color Evolution of Sources with AGN spectral features}
\label{sec:appendix_ced}

\begin{figure}
\includegraphics[width=1\linewidth]{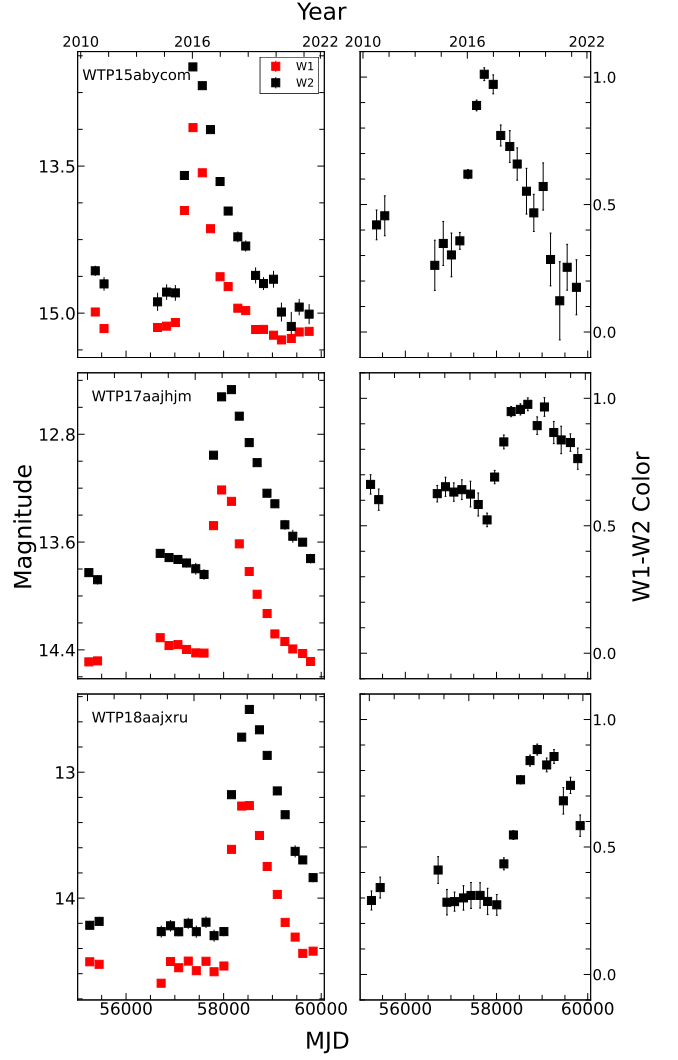}
    \caption{Color evolution diagrams for sources that were noted as displaying inconclusive color evolution. These sources were later eliminated owing to their spectra displaying features characteristic of AGN.}
\label{fig:eliminated_ced}
\end{figure}

As noted in Section \ref{sec:photo_analysis}, we concluded that the objects WTP15abycom, WTP17aajhjm and WTP18aajxru showed inconclusive color evolution, displaying  variations in the flux light curve and color evolution that were more characteristic of AGN flares. The color evolution diagrams for these sources are displayed in Figure \ref{fig:eliminated_ced}, as are the aperture photometry light curves. WTP15abycom becomes increasingly bluer as the brightness of the flare decreases: its post-flare color is clearly lower than its pre-flare color, which is not expected for a non active TDE host galaxy. WTP17aajhjm and WTP18aajxru both show slight variability in color evolution prior to the start of the flare, while the former also turns bluer in the beginning of the flare. Additionally, WTP17aajhjm shows pre-flare variation in its flux light curve. While we initially opted to keep these sources in the sample out of an abundance of caution, they were found to exhibit AGN spectral features in our subsequent analysis and were eventually excluded from the final sample of TDEs. This furthers validates the strength of the proposed color based criterion to select IR-bright TDEs.

\section{Dust Modeling} \label{sec:appendix_dust}

\begin{figure}[]
    \centering
    \includegraphics[width=1\linewidth]{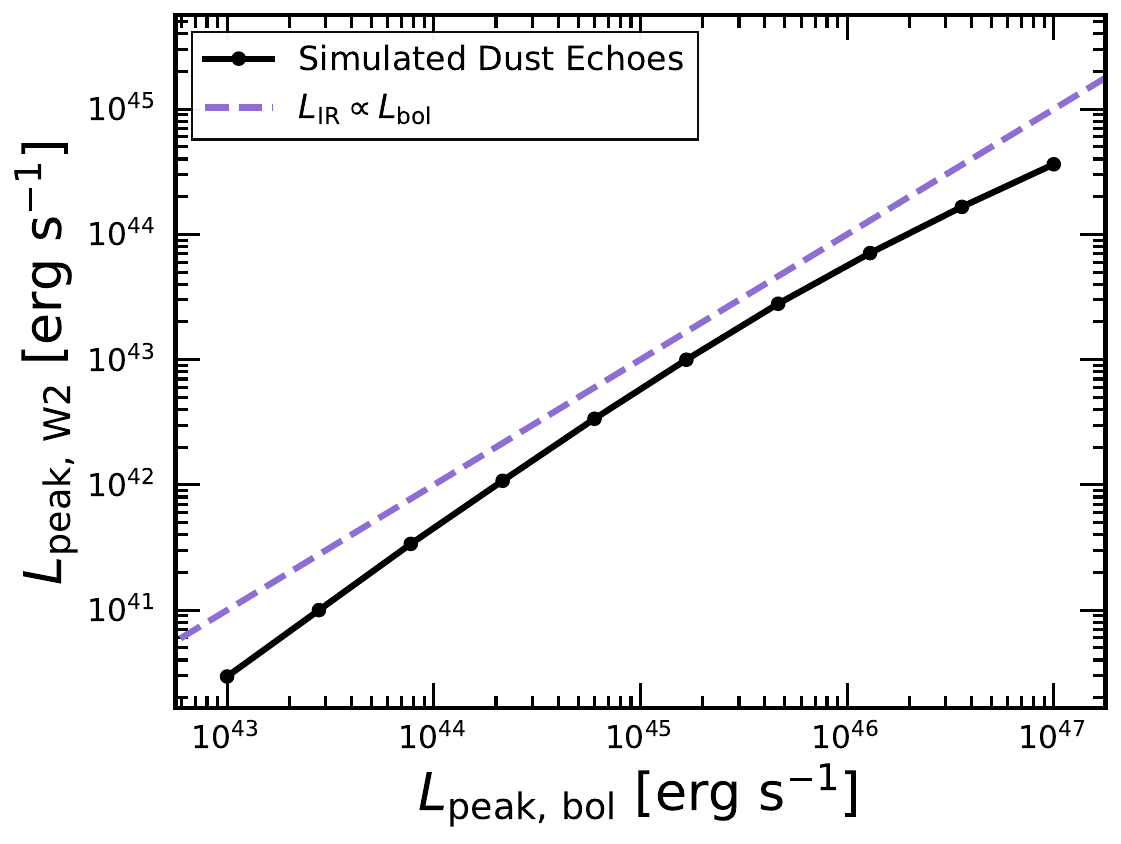}
    \caption{Comparison of the peak W2 luminosity to the input peak bolometric luminosity for a simple dust shell that is optically thin to its own emission. This assumes MRN dust with a single grain size of $a = 0.1~\mu$m and a density profile that scales like $r^{-1}$. The sublimation radius is depends on the input peak bolometric luminosity like $r_\mathrm{sub} \propto L_\mathrm{bol}^{1/2}$. We do not see a cut-off in the observed WISE W2 luminosity based on the effects of dust sublimation.}
    \label{fig:ir-to-bol}
\end{figure}

As discussed in Section \ref{sec:disc_cutoff}, we used the time-dependent dust emission model from \cite{Masterson2025ApJ} to test the effects of dust sublimation on the observed peak W2 luminosity. Figure \ref{fig:ir-to-bol} shows that there is no clear cut-off in the peak IR luminosity. This figure was produced by leaving the properties of the nuclear dust shell fixed, with the exception of the peak bolometric luminosity of the input flare. This model assumes that no emission comes from within the sublimation radius set by this peak bolometric flux, given by $r_\mathrm{sub} \approx 0.15 (L_\mathrm{bol} / 10^{45}\, \mathrm{erg\, s}^{-1})$~pc. The figure shows that there is no strong cut-off from the effects of smearing out the integrated light over a larger radius in more luminous flares, thereby confirming that the turn-over we see in our luminosity function is likely intrinsic to the TDE event horizon suppression effect.

\section{Additional Information on Optical Spectra} \label{sec:appendix_spectra}

\begin{figure*}

\includegraphics[width=1.0\linewidth]{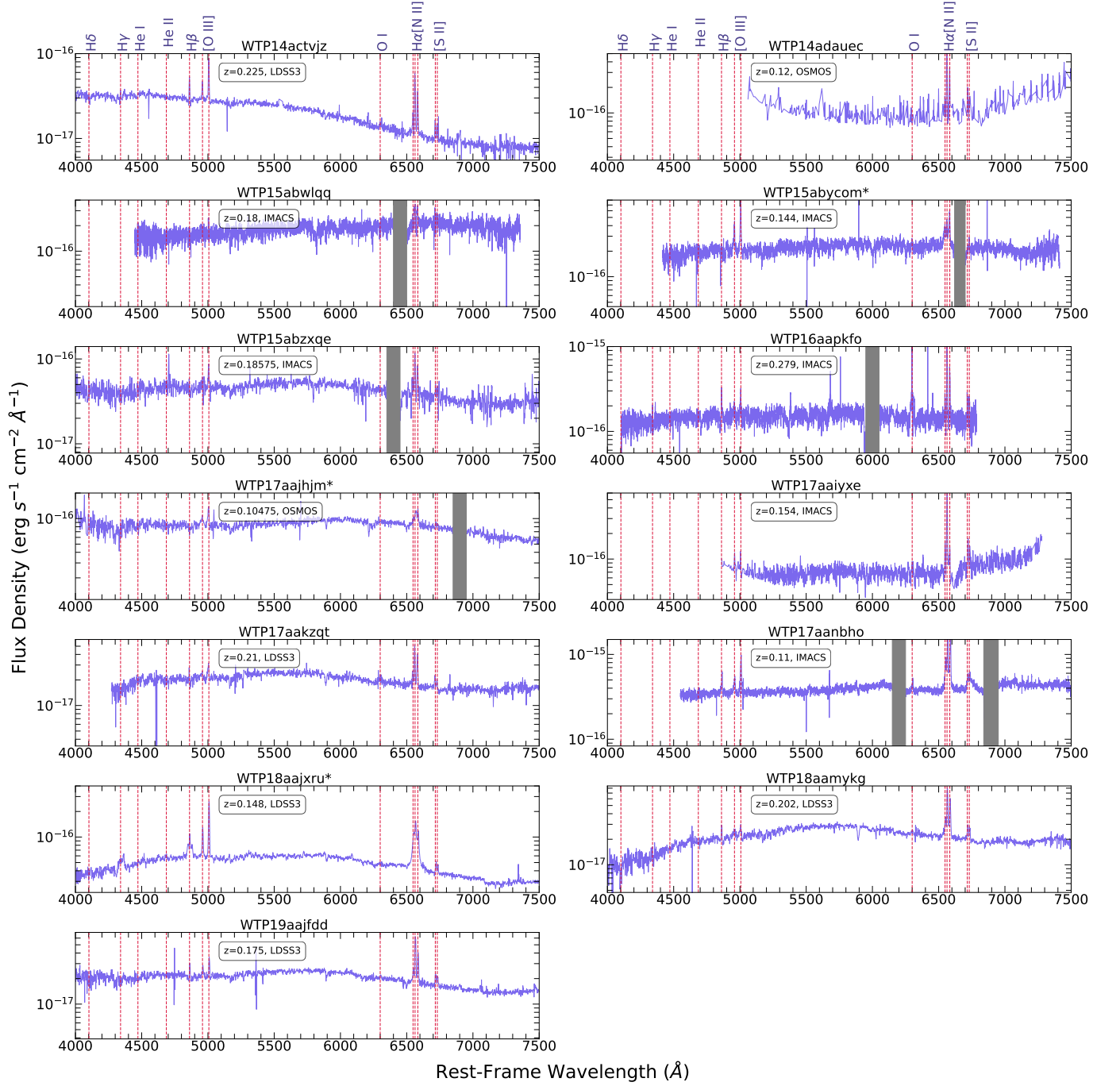}
    \caption{The spectra  collected for this study. The red dashed lines represent relevant emission lines, labeled at the top of the figure. Grey shaded regions represent telluric regions of absorption. The sources displayed with an asterisk next to their name, WTP15abycom, WTP17aajhjm and WTP18aajxru, exhibit AGN spectral features and were thus excluded from the final sample.}
\label{fig:all_spectra}
\end{figure*}

\begin{table*}

  \caption{Observation details on the optical spectra collected for this work. All spectra were collected well after the detection of their corresponding IR flare. }
  \centering
  \resizebox{1.1\textwidth}{!}{
  \begin{tblr}{
      colspec={lccccccc},
      row{1}={font=\bfseries},
      row{even}={bg=white!10}}
      \hline
      \hline     
            \text{WTP Name} & \text{Date} & \text{Telescope} &\text{Instrument}& \text{Grating/Grism}& \text{Slit Width (")} &\text{Wave Range(A)} &\text{Exposure Time(s)}\\
    \hline
 
    WTP14actvjz&2024-01-09&Magellan/Clay&LDSS3&VPH-All&0.75&4250-10000&2700\\
    WTP14adauec& 2025-03-16&MDM Hiltner&OSMOS&VPH-Red&1.0 Outer&5000-8500&3600\\
    WTP15abwlqq&2024-09-12&Magellan/Baade&IMACS&Gra-600-8.6&0.7&5250-8680&3600\\ 
    WTP15abycom&2024-09-12&Magellan/Baade&IMACS&Gra-600-8.6&0.7&5050-8480&3600\\
    WTP15abzxqe&2024-09-12&Magellan/Baade&IMACS&Gra-600-8.6&0.7&3780-10580&3600\\ 
    WTP16aapkfo&2024-09-12&Magellan/Baade&IMACS&Gra-600-8.6&0.7&5250-8680&3600\\ 
    WTP17aaiyxe&2025-02-24&MDM Hiltner&OSMOS&VPH-Red&1.0 Outer&5000-8500&3600\\
    WTP17aajhjm&2024-09-12&Magellan/Baade&IMACS&Gra-600-8.6&0.7&4250-10670&1500\\
    WTP17aakzqt&2024-01-09&Magellan/Clay&LDSS3&VPH-All&0.75&4250-10000&1800	\\ 
    WTP17aanbho&2024-09-12&Magellan/Baade&IMACS&Gra-600-8.6&0.7&5050-8480&3600\\ 
    WTP18aajxru&2024-01-09&Magellan/Clay&LDSS3&VPH-All&0.75&4250-10000&2700\\ 
    WTP18aamykg&2024-01-09&Magellan/Clay&LDSS3&VPH-All&0.75&4250-10000&2700\\
    WTP19aajfdd&2024-01-09&Magellan/Clay&LDSS3&VPH-All&0.75&4250-10000&2700\\
    \hline
\end{tblr}    
}
\label{tab:opt_spectra}
\end{table*}

We present the reduced optical spectra collected for each object in our final sample in Figure \ref{fig:all_spectra}. Further details on instruments used and the dates the spectra were taken can be found in Table \ref{tab:opt_spectra}. All spectra in our sample were obtained at least 4 years after the IR flare took place, as can be concluded from the observing dates provided in Table \ref{tab:opt_spectra}.  As can be seen in Figure \ref{fig:all_spectra}, all except for 2 objects display narrow emission lines, which is expected for TDEs at late times. Broad emission is often only observed within roughly the first year after the peak flux \citep[e.g.][]{nicholl_tidal_2019}. 

The spectrum for WTP18aajxru was obtained using the Low Dispersion Survey Spectrograph (LDSS-3) on the Magellan/Clay telescope on 2024-01-09. This spectrum was flagged as having broad H-$\beta$ and H$\alpha$ emission lines, which we defined in Section \ref{sec:spec_analysis} as having a line width $\sigma$ of $\geq$ \SI{500}{\km\per\s}. Therefore, WTP18aajxru was classified as an AGN and was excluded from our sample for further analysis.

The spectrum for WTP17aajhjm was obtained using the Magellan-Baade telescope's Inamori Magellan Areal Camera and Spectrograph (IMACS) on 2024-09-12. This spectrum was also flagged as having broad H$\alpha$ emission. Combined with the variation it displayed in color evolution, as detailed in Section \ref{sec:photo_analysis}, we deemed these features as characteristic of AGN flares and excluded WTP17aajhjm from the remainder of the analysis.

The spectrum for WTP15abycom was also obtained using Magellan-Baade's IMACS instrument on 2024-09-12. 

Its spectrum shows narrow emission lines, similar to our other TDE candidates. However, following the BPT analysis outlined in Section \ref{sec:spec_analysis}, WTP15abycom was classified it as an AGN, mostly due to the prominent [N II] emission line, which is stronger than the H$\alpha$ line (Fig. \ref{fig:all_spectra}). Hence, this object was subsequently excluded from our final sample. 

As noted in Section \ref{sec:spec_analysis}, the spectra we obtained for WTP14adauec and WTP17aaiyxe, using the Ohio State Multi-Object Spectrograph(OSMOS) on the MDM Hiltner telescope, were not of a sufficiently high quality to run the pPXF fitting. However, as can be seen from their raw spectra, they do not display broad hydrogen lines or other spectral features characteristic of AGN, and so we included them in our final sample of TDEs.

\bibliography{tde_wise}{}
\bibliographystyle{aasjournal}

\end{document}